\documentclass[aps,prb,superscriptaddress,showpacs,twocolumn,a4paper]{revtex4}
\usepackage{epsfig}
\usepackage{amsmath}

\normalsize

\def\bj{{\bf j}}
\def\bk{{\bf k}}

\def\b0{{\bf 0}}

\def\up{\uparrow}
\def\down{\downarrow}

\def\eps{\epsilon}

\def\Gam{\Gamma}

\def\Lam{\Lambda}

\def\sg{\sigma}
\def\Sg{\Sigma}

\def\mb#1{{\mathbf #1}}

\begin{document}

%%%%%%%%%%%%%%%%%%%%%%%%%%%%%%%%% TITLE PAGE %%%%%%%%%%%%%%%%%%%%%%%%%%%%

\title{Strong correlations and formation of ``hot spots''\\
in the quasi-one-dimensional Hubbard model at weak coupling}

\author{Daniel Rohe}
\affiliation{Centre de Physique Th\'eorique, Ecole Polytechnique, CNRS-UMR 7644,
91128 Palaiseau Cedex, France}

\author{Antoine Georges}
\affiliation{Centre de Physique Th\'eorique, Ecole Polytechnique, CNRS-UMR 7644,
91128 Palaiseau Cedex, France}

\date{\small\today}

%%%%%%%%%%%%%%%%%%%%%%%%%%%%%%%%% ABSTRACT %%%%%%%%%%%%%%%%%%%%%%%%%%%%%%

\begin{abstract}

We study the anisotropic two-dimensional Hubbard model at and near half filling
within a functional renormalization group method, focusing on the structure of
momentum-dependent correlations which grow strongly upon approaching a critical
temperature from above. We find that a finite nearest-neighbor interchain
hopping is not sufficient to introduce a substantial momentum dependence of
single-particle properties along the Fermi surface. However, when a
sufficiently large second-nearest neighbor inter-chain hopping is introduced,
the system is frustrated and we observe the appearance of so-called ``hot spots'',
specific points on the Fermi surface around which scattering becomes particularly
strong. We compare our results with other studies on quasi-one-dimensional systems.

\end{abstract}

\pacs{71.10.-w, 71.27.+a, 71.20.Rv, 78.30.Jw}

\maketitle

%%%%%%%%%%%%%%%%%%%%%%%%%%%%%%%%% PAPER %%%%%%%%%%%%%%%%%%%%%%%%%%%%%%%%

\section{Introduction}

The one-band Hubbard model serves as a minimal model for various correlated electron systems, since it is
capable of capturing a number of non-trivial phenomena which are due to the interplay between kinetic and
potential energy. In one dimension, numerous theoretical methods are available which have led to a thorough
understanding of the low-energy physics.\cite{Gia_book} In higher dimensions, however, rigorous statements are
scarce and many controversies remain. It is therefore natural to ask the question how the cross-over from one to
two or higher dimensions takes place. Furthermore, since the discovery of quasi-one-dimensional organic
conductors and superconductors we have access to materials which are realizations of this physical situation. In
these compounds many interesting observations were made during the last two decades, some of which are still
calling for a conclusive theoretical description.\cite{Bou_99}

In this work we
will consider the evolution of a model system near
half-filling, upon increasing the dimensionality via an increase in the perpendicular kinetic-energy coupling between
one-dimensional chains. To tackle this question, we employ a functional renormalization group (fRG)
technique, which provides a rigorous framework for the computation of low-energy properties starting from a
microscopic model.\cite{Sal_book} While this method reduces to the well-known g-ology RG in one dimension, it
has been applied successfully to the two-dimensional case, where the angular dependence of the coupling function
along the Fermi surface needs to be taken into account.\cite{Hal_00} By allowing for anisotropic hopping
parameters we are in principle able to access the complete region from the one-dimensional to the
two-dimensional case. In this work we will fix the degree of anisotropy and study the behavior of the system as
a function of other parameters.

%Status

The dimensional cross-over in quasi-one-dimensional systems (and the intimately related phenomenon of
``deconfinement'', i.e the Mott insulator to metal transition induced by increasing the inter-chain kinetic
energy) have been investigated in recent years within several theoretical approaches. There is consensus on some
aspects, but some disagreements between these studies do remain (mainly due to the different theoretical tools
which have been employed, and the different regimes of parameters which have been investigated).
The issue as such was raised by experimental studies on Beechgard salts. In
particular, the optical spectroscopy experiments of Vescoli et al. revealed an insulator-to-metal transition as
a function of increasing interchain hopping parameter, which changes with chemical composition~\cite{Ves_98}. At
the same time, Bourbonnais and J\'{e}rome discussed these results within a scenario where a one-dimensional Mott
insulator evolves into a metallic state when the inter-chain hopping reaches the order of the Mott
gap.\cite{Bou_98}.

Subsequently, several model calculations where made to substantiate and verify this concept.
Biermann et al. employed an extension of dynamical mean-field theory (DMFT), the so-called chain-DMFT, which
replaces the original problem by a chain self-consistently coupled to a bath, while taking into account the full
intra-chain momentum dependence ([\onlinecite{Bie_01}], see also [\onlinecite{Arr_99}]).
They do indeed find a transition from an insulating to a metallic
state when the interchain hopping is increased, as well as a crossover from a Luttinger liquid to a Fermi
liquid at fixed interchain hopping, when the temperature is decreased.
%
%An important thing to note in these
%results is that in the metallic state the quasi-particle properties vary only very slightly along the resulting
%quasi-1d Fermi surface, and special points with exceptionally large scattering rates or small quasi-particle
%weights are absent.
%
Essler and Tsvelik considered this problem starting from a one-dimensional Mott insulator and using a
resummed expansion in the inter-chain hopping~\cite{Ess_02,Ess_05}. Using this approach, they
suggested that the metallic phase does not develop immediately with a large Fermi surface
resembling the non-interacting one. In contrast, close enough to the Mott insulating phase,
Fermi surface pockets appear in specific locations, while large parts of the
would-be Fermi surface remain gapped due to the influence of the one-dimensional Mott gap. Since they use a
particular dispersion in the interchain direction which is particle-hole symmetric, they eventually find that
this pocket Fermi liquid becomes unstable towards an ordered state, however at much lower temperatures as
compared to the Mott gap. The location of these pockets
%at $k_b=0$ ($k_b=\pi$), slightly above (below) the one-dimensional Fermi wave vector
%$k_a = \pi/2$, where $k_a$ refers to the momentum in chain direction and $k_b$ to the momentum
%perpendicular to the chains.
%
is such that the neighbourhood of the point $k_a=k_b=\pi/2$ (with $k_a$ the momentum along the chain direction and
$k_b$ perpendicular to it) is gapped out and not part of the Fermi surface.
This point thus corresponds to a ``hot spot'', at which the scattering rate is very large (and can
even lead to a complete suppression of quasiparticles at this point).
While the chain-DMFT studies of Ref.~[\onlinecite{Bie_01}] did not observe this phenomenon (perhaps because of the
range of coupling or temperature), more recent studies of an anisotropic spinless model using chain-DMFT
did observe a partial destruction of the Fermi surface with hot spots at the same
location~\cite{Ber_06}.

From the weak-coupling standpoint, early calculations suggested the occurrence
of hot spots from a simple perturbative calculations of the scattering rate\cite{Zhe_95}.
Within a renormalization group treatment, Duprat and Bourbonnais found that for a finite and fixed value of the
interchain hopping the influence of strong spin-density-wave fluctuations can lead to anisotropic scattering
rates along a quasi-one-dimensional Fermi surface, leading to the emergence of hot spots\cite{Dup_01}. However,
the locations of cold and hot regions are exactly exchanged with respect to the findings by Essler and Tsvelik, with
the hot spots found at $k_b=0$ and $k_b=\pm\pi$ in Ref.~[\onlinecite{Dup_01}].
It should be noted that these results were obtained for a system away from half filling, meaning that umklapp
processes are suppressed in the RG treatment. In the present work, we focus on the half-filled case or its
immediate vicinity, and use a functional RG technique which does take
umklapp processes into account.

Thus, there exists a consensus that ``hot spots'' might form in quasi one-dimensional systems, but the
various treatments do not agree on their location. It is tempting to speculate that this simply
reflects the different locations of these hot regions in the weak and strong coupling limits. At any rate,
there are some compelling experimental indications for strongly anisotropic scattering rates in
quasi one-dimensional organic conductors, as pointed out early on in Ref.~[\onlinecite{Cha_92}] and dicussed
further in the conclusion.

% Model

\section{Model}

We consider the one-band Hubbard model:
\begin{equation}
 H = \sum_{\bj,\bj'} \sum_{\sg}
 t_{\bj\bj'} \, c^{\dag}_{\bj\sg} c^{\phantom{\dag}}_{\bj'\sg} +
 U \sum_{\bj} n_{\bj\up} n_{\bj\down}
\end{equation}
with a local repulsion $U>0$ and hopping amplitudes $t_{\bj\bj'} = -t_a$ between nearest neighbors in the
a-direction along the chains, $t_{\bj\bj'} = -t_b$ between nearest neighbors in the b-direction perpendicular to
the chains, and $t_{\bj\bj'} = -t_{b}^{'}$ between second-nearest neighbors in the b-direction. The
corresponding dispersion relation reads

\begin{equation}
 \eps_{\bk} = -2t_a\cos k_a -2t_b \cos k_b - 2t_{b}^{'} \cos 2k_b .
\end{equation}
At half-filling and $t_{b}^{'}=0$ the non-interacting Fermi surface is perfectly nested. The introduction of a
finite $t_{b}^{'}$ will allow us to study the effects of deviation from this perfect nesting condition. It is
important to note that we do \emph{not} linearise the dispersion in the chain direction, as is commonly done in
RG treatments originating from the 1d g-ology setup.\cite{Dup_01} Therefore, the Fermi surface is perfectly
nested \emph{only} at half filling. Away from half filling perfect nesting is destroyed, even without a finite
value of $t_{b}^{'}$. In the following all energies are given in terms of $t_a$ which we set to unity.

%%% Method

\section{Method}

\begin{figure}
\center \epsfig{file=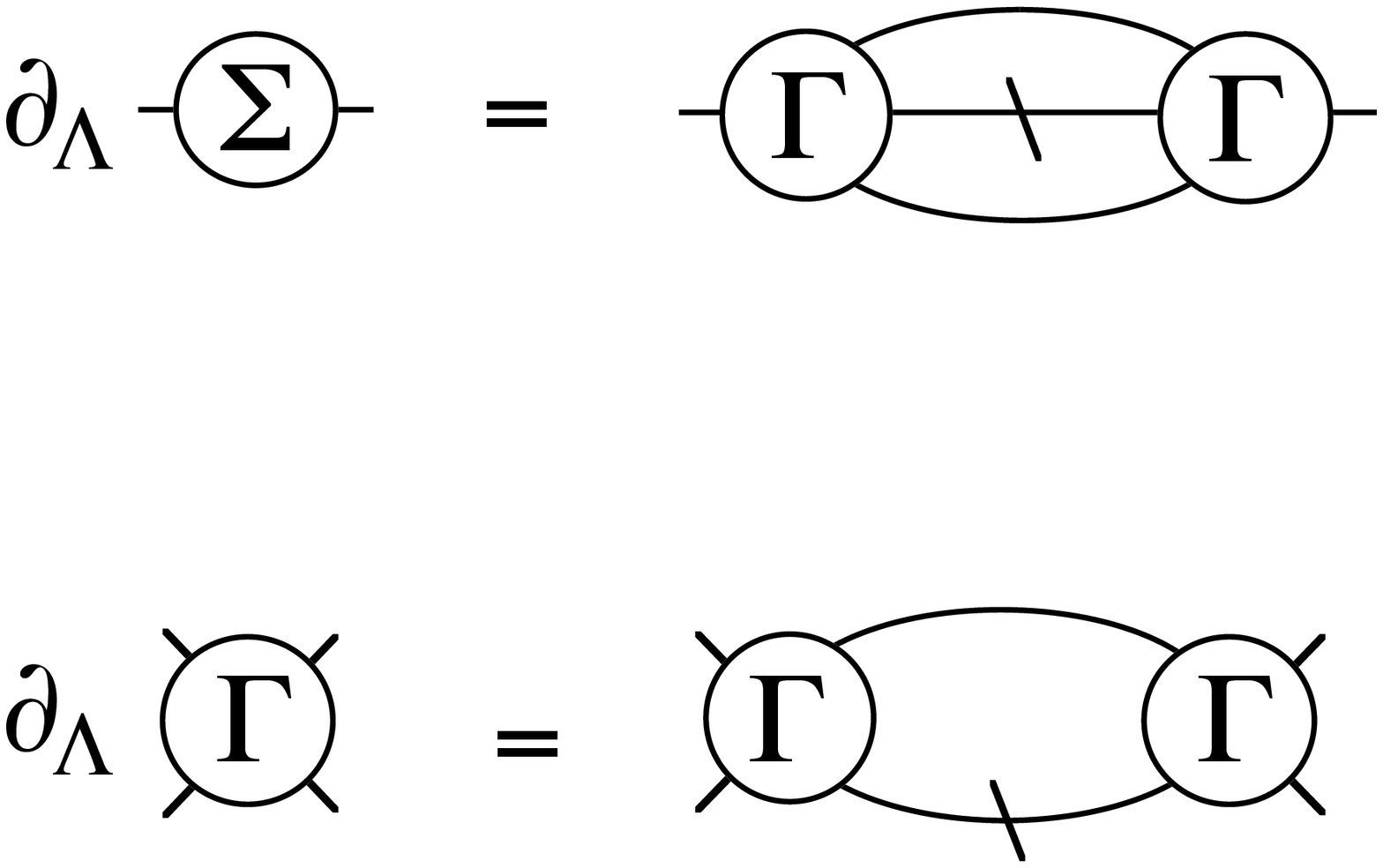,width=9cm} \caption{Flow equations for the self energy $\Sg^{\Lam}$ and
 the two-particle vertex $\Gam^{\Lam}$, respectively;
 the internal lines without slash correspond to the bare propagator
 $D^{\Lam}$, the lines with a slash to its $\Lam$-derivative
 $\partial_{\Lam} D^{\Lam}$.}
\label{fig1.eps}
\end{figure}

\begin{figure}
\center \vskip 2cm \epsfig{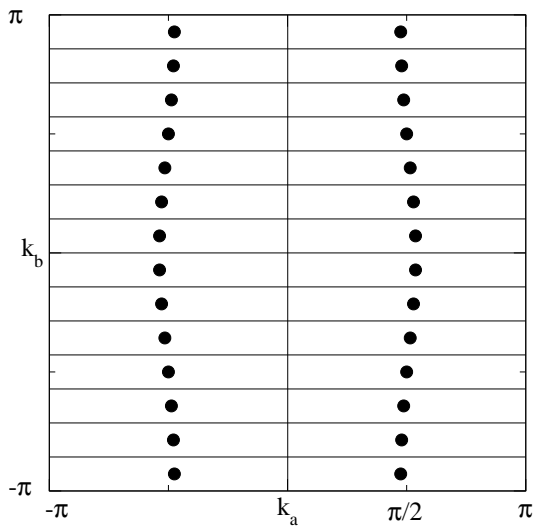} \caption{Patching scheme for $N=28$.} \label{7.eps}
\end{figure}

In direct analogy to the technique applied in reference [\onlinecite{Roh_05}] we use the Wick-ordered version of
the fRG to compute the one-loop flow of the interaction vertex and the two-loop flow of the self energy, as
depicted in Fig.\ 1. The internal lines without slash in the Feynman diagrams correspond to the bare propagator
\begin{equation}
 D^{\Lam}(k) = \frac{\Theta(\Lam - |\xi_{\bk}|)}
 {ik_0 - \xi_{\bk}} \; ,
\end{equation}
where $\xi_{\bk} = \eps_{\bk} - \mu$ and $\Lam > 0$ is the cutoff; the lines with slash correspond to
$\partial_{\Lam} D^{\Lam}$, which is proportional to $\delta(\Lam - |\xi_{\bk}|)$.

We parametrize the interaction vertex $\Gamma$ by its static values on a reduced number of points/patches on the
non-interacting Fermi surface, as illustrated in Fig.\ 2. It is thus parametrised by a momentum-dependent
singlet (triplet) component $\Gamma_{s (t)}(k_1,k_2,k_3)$, where the $k_i$ constitute a discrete set of momenta
on the Fermi surface. We stress that this does not correspond to treating a finite system, since internal
integrations are done in the thermodynamic limit.

In the present work, we do not directly compute the flow of the self-energy, but rather infer from
the properties of the two-loop diagram what the influence of a strongly renormalized vertex function on the self
energy will be. In the case of the two-dimensional Hubbard model this calculation has been done
explicitly,\cite{Roh_05} giving us confidence with respect to this reasoning.

\section{Results}

\subsection{$t_{b}^{'}=0$ - perfect nesting}

\begin{figure}
\center \vskip 2cm \epsfig{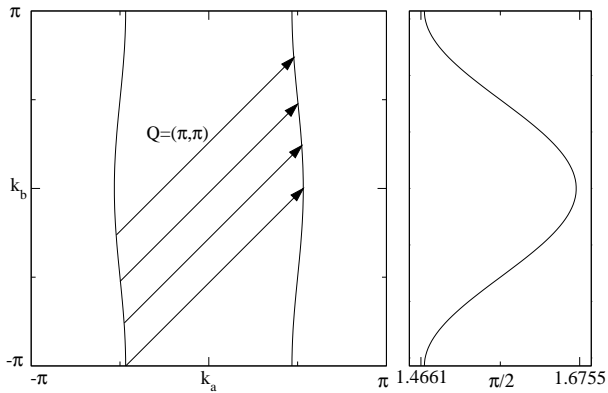} \caption{Fermi surface and nesting vector for $t_b=0.1$ and
$t_{b}^{'}=0$, that is the case of perfect nesting (All energies in units of $t_a=1$). The picture on the right
shows the warping of the Fermi surface near $k_a=\frac{\pi}{2}$.} \label{fspn.eps}
\end{figure}

\begin{figure}
\center \vskip 2cm \epsfig{file=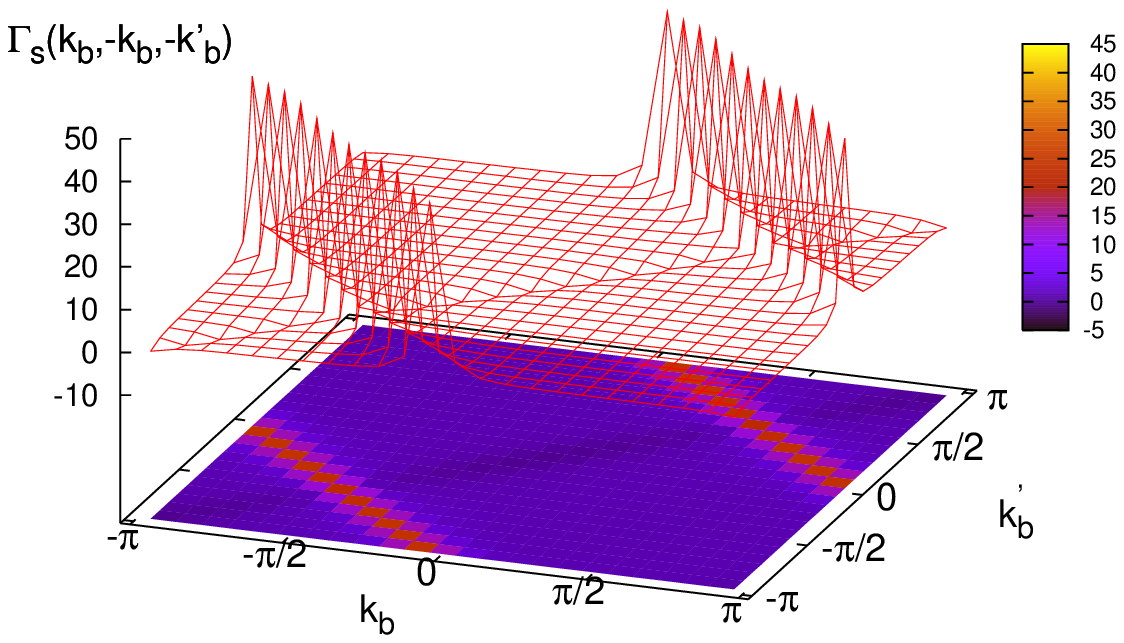,width=7.5cm} \epsfig{file=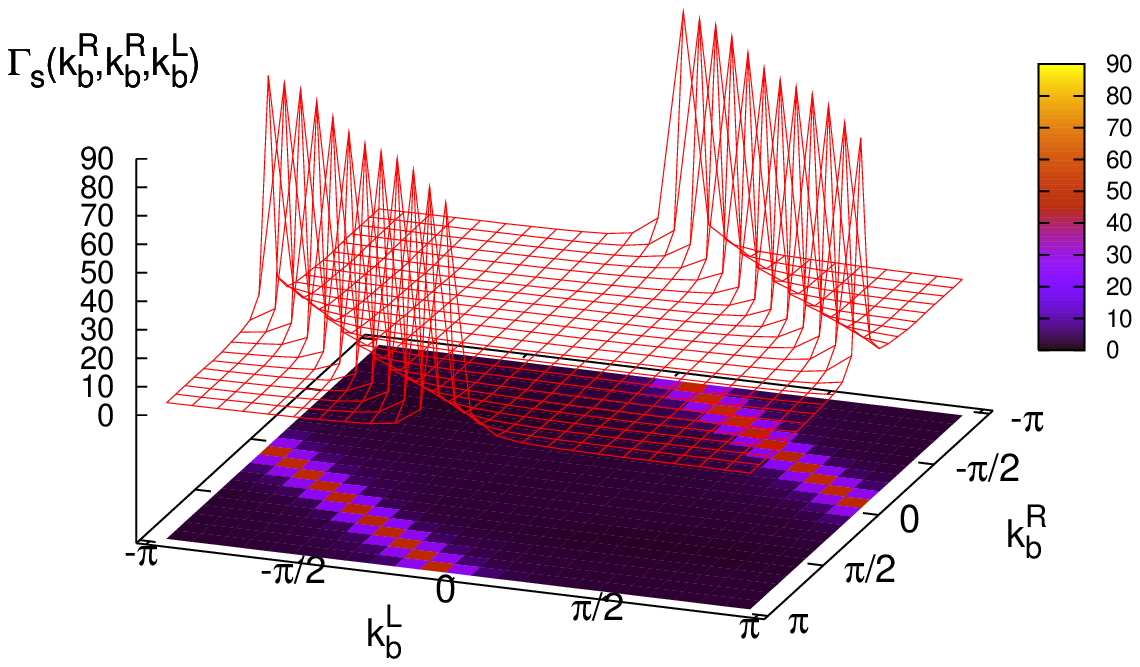,width=7.5cm}
\caption{\emph{Left}: Singlet Vertex in the Cooper channel on the Fermi surface as a function of $k_b$ and
$k'_b$. \emph{Right}: Singlet Vertex in the Twin-Umklapp channel on the Fermi surface as a function of $k^R_b$
and $k^L_b$. Here $t_b=0.1$, $t_b^{'}=0$, $T=0.011$, $\mu=0$, $U=2.45$ (All energies in units of $t_a=1$). In
this case the Fermi surface is essentially perfectly nested.} \label{nested}
\end{figure}

At $t_{b}^{'}=0$ and $\mu=0$ the non-interacting Fermi surface is perfectly nested with nesting vector
$\mb{Q}=(\pi,\pi)$ and defines the so-called Umklapp surface as illustrated in Fig.\ 3. For $t_{b}=0$ the
problem reduces to the one-dimensional half-filled Hubbard model. In this case the divergence of Umklapp
couplings at low energies signals the onset of the Mott insulating phase.\cite{Bou_04} When we analyze the
one-loop flow of the interaction vertex for finite values of the interchain hopping $t_{b}$, we find that the
divergence of Umklapp processes persists at all finite values of $t_{b}$, which we chose to be $t_b=0.1t_a$
throughout this work. Namely the one-dimensional Umklapp couplings connect to two-dimensional Umklapp processes
of the type $(k_F,k_F^\prime) \rightarrow (k_F+{\mb Q},k_F^\prime-{\mb Q})$ with momentum transfer $\mb{Q} = (\pi,\pi)$. The
crucial point is that this divergence is nearly perfectly homogeneous along the Fermi surface. Due to the
feedback of the interaction onto the self energy via the two-loop diagram, this implies that there are no
isolated hot spots at which the dominant scattering processes are dominant compared to other regions on the
Fermi surface. Instead, {\it the whole Fermi surface} is ''hot''. Along with Umklapp processes the interaction
develops divergences in the Cooper channel, also owing to the importance to scattering with wave vector
$\mb{Q}=(\pi,\pi)$. The behavior of the coupling function illustrates this very clearly. In the left plot of
figure \ref{nested} we display the singlet component of the interaction function $\Gamma_S(\mb{k},-\mb{k},-\mb{k}')$ on
the Fermi surface as a function of $k_b$ and $k'_b$ at the end of the flow, in direct analogy to the analysis
presented in reference [\onlinecite{Dup_01}]. The bare interaction is chosen in such a way that the temperature is
slightly above the pairing temperature, at which the flow of the vertex function diverges. The interaction is
homogeneously peaked along the lines $k'_b = \pi - k_b$ and  $k'_b = -\pi - k_b$. We name these lines ''Peierls
lines'' since they correspond to scattering processes in which a momentum $\mb{Q}$ is exchanged between the two
incoming particles and $\mb{Q}$ is the generalization of $2k_F$ in one dimension at half filling and perfect
nesting.

We thus see that \emph{all} points on the Fermi surface are equally strongly affected by strong correlations
appearing in the Cooper channel, which may eventually lead to a phase where \emph{all} one-particle states along
the Fermi surface exhibit a pseudogap. Since we look at a system at half filling it is essential to consider
Umklapp processes, as mentioned above. We therefore extend the analysis offered in reference [\onlinecite{Dup_01}]
and show in figure \ref{nested} in the right plot the interaction function $\Gamma_S(k_{b}^R,k_{b}^R,k_{b}^L)$
on the Fermi surface in the so-called twin-Umklapp channel, meaning that both incoming momenta are identical,
and $k_{b}^R (k_{b}^L)$ are momenta corresponding to right (left) movers in the standard terminology familiar
from the one-dimensional case. We see that along the lines $k_{b}^L = k_{b}^R - \pi$ and $k_{b}^L = k_{b}^R +
\pi$ the coupling function in this channel is also homogeneously peaked, confirming the conclusion drawn on the
basis of the Cooper channel. We recall that in contrast to the one-dimensional case the system will eventually
undergo a transition into an antiferromagnetic state at zero temperature. Here, however, we are concerned with
finite-temperature precursor effects, which may legitimately be compared.

These results are distinct from those obtained in other RG calculations\cite{Dup_01}. There, perfect nesting is
artificially introduced due to a linearization of the dispersion in the chain direction, and Umklapp processes
are neglected. The divergent couplings are then found only in the Peierls section of the Cooper channel for
sufficiently small $t_{b}^{'}$, leading to isolated hot spots at $k_b=0$ and $k_b=\pm\pi$.

\subsection{$t_{b}^{'}\neq0$ - effects of frustration}

\begin{figure}
\center \vskip 2cm \epsfig{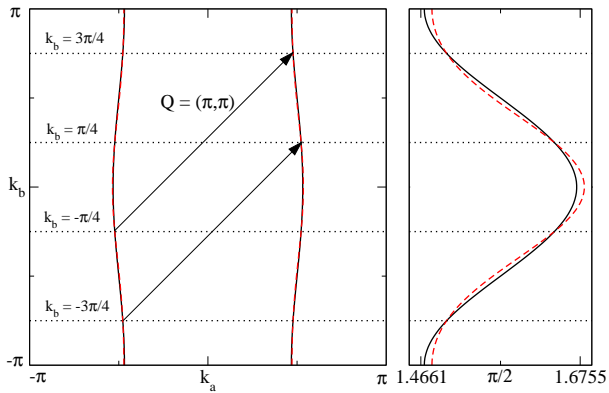} \caption{Fermi surface for $t_b=0.1$ and $t_{b}^{'}=0.1t_b$
(red, dotted line) and for $t_{b}^{'}=0$ (black line). The points at which the two intersect allow for Umklapp
scattering of the type $(k_F,k_F) \rightarrow (k_F+{\mb Q},k_F-{\mb Q})$ at arbitarily low energies and
therefore emerge as hot spots in the RG flow.} \label{fsnn.eps}
\end{figure}

\subsubsection{$\mu=0$ - half filling}

\begin{figure}
\center \vskip 2cm \epsfig{file=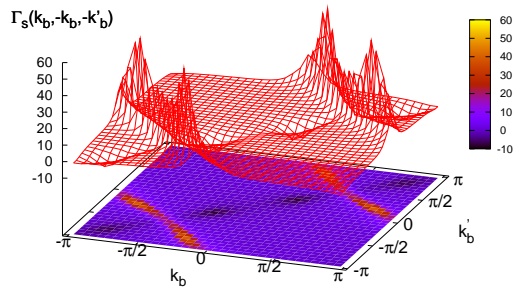,width=7.5cm} \hfill
\epsfig{file=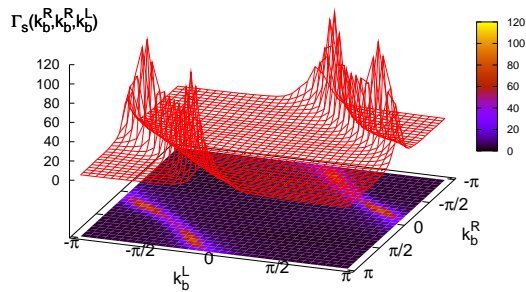,width=7.5cm} \caption{Same as figure \ref{nested}, but here $t=0.1$,
$t'=0.1t_b$, $T=0.011$, $\mu=0$, $U=3.08$. In this case perfect nesting is destroyed and the system is
frustrated.} \label{frustrated}
\end{figure}

A finite second-nearest neighbor hopping $t_{b}^{'}$ will destroy the nesting condition for all wave vectors on
the Fermi surface, except for a few special points. We set the chemical potential to $\mu=0$ and use $t_b=0.1$
and $t_{b}^{'}=0.1t_b$. In this case the non-interacting system essentially remains half filled and the
frustrated Fermi surface intersects the Umklapp surface at $k_b=\pm\pi/4$ and $k_b=\pm3\pi/4$, as shown in Fig.
\ref{fsnn.eps}. Then, at arbitrarily low energies Umklapp scattering of the type $(k_F,k_F) \rightarrow
(k_F+{\mb Q},k_F-{\mb Q})$ is possible if and only if $k_F$ is located at the intersection between the
non-nested Fermi surface and the Umklapp surface. The resulting RG flow of the interaction vertex shows a
dominant divergence of the couplings corresponding to exactly these processes. This can best be seen in figure
\ref{frustrated}, where the coupling function is shown in analogy to figure \ref{nested}. In contrast to the
case $t'=0$ the coupling function in both channels shows a strongly peaked behavior which is not homogeneous
along the Peierls lines, but which is peaked at the points where the Fermi surface intersects the Umklapp
surface. This is reminiscent of a scenario in which there exist so-called hot spots, that is special points at
which the scattering rate is particularly large or a pseudogap may appear in the spectral function, in analogy
to the case of a two-dimensional system.\cite{Roh_05} Note that for the frustrated system the Peierls lines
defined above in the plots of the vertex function correspond to scattering processes with wave vector $\mb{Q}$
\emph{only} at hot spots. Elsewhere the momentum transfer on the Fermi surface is incommensurate.

\subsubsection{$\mu \neq 0$ - slightly doped system}

\begin{figure}
\center \vskip 2cm \epsfig{file=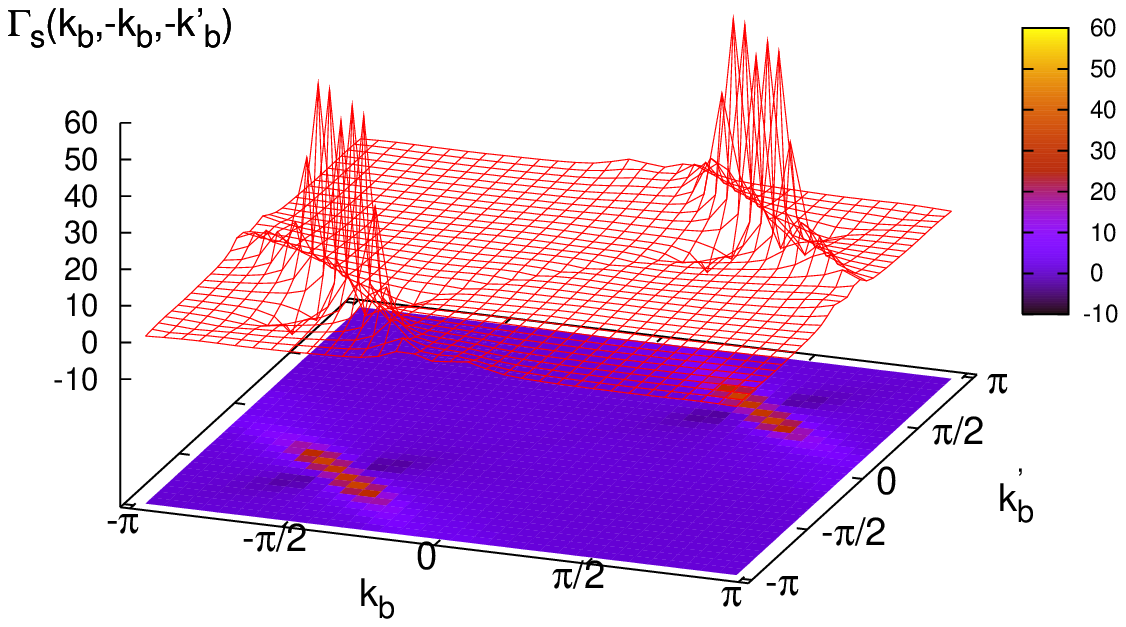,width=7.5cm} \epsfig{file=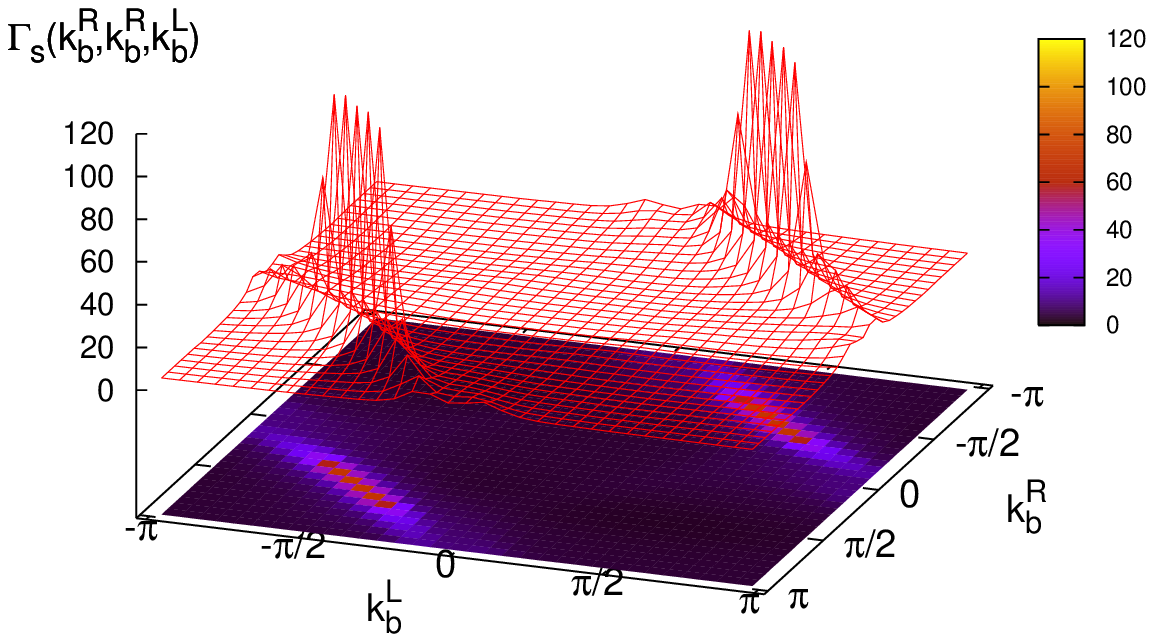,width=7.5cm}
\caption{Same as Figure \ref{frustrated}, but here $\mu=0.02$, $U=3.126$. In this case the Fermi surface touches
the Umklapp surface at $(\pi/2,\pi/2)$.} \label{edp}
\end{figure}

\begin{figure}
\center \vskip 2cm \epsfig{file=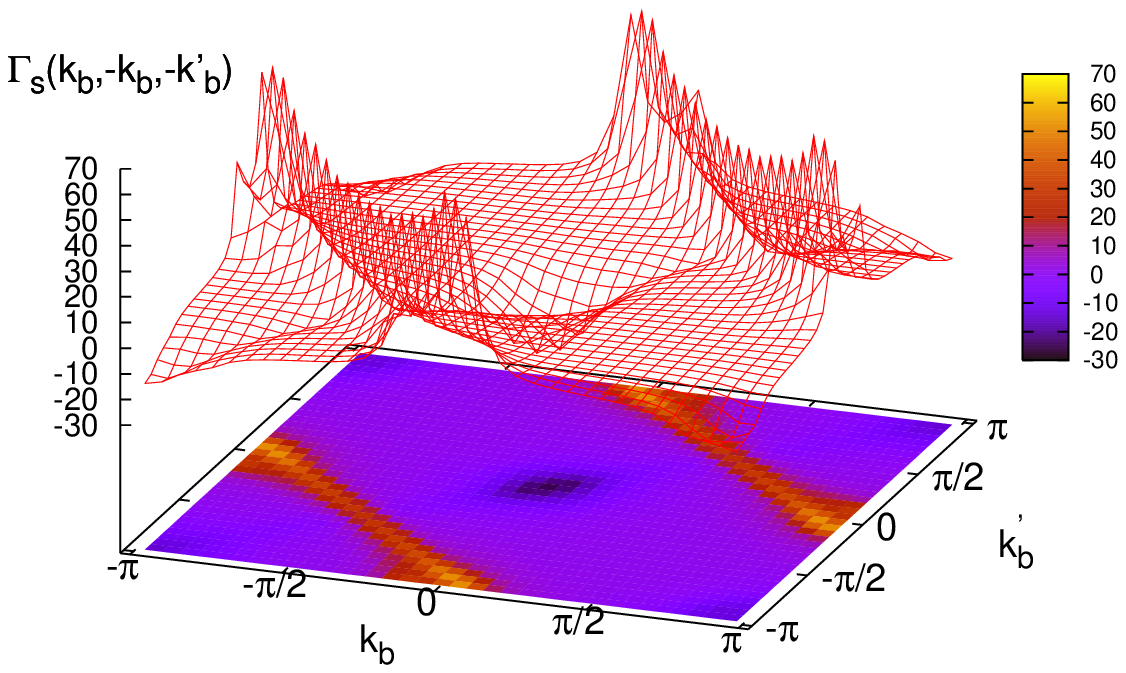,width=7.5cm} \epsfig{file=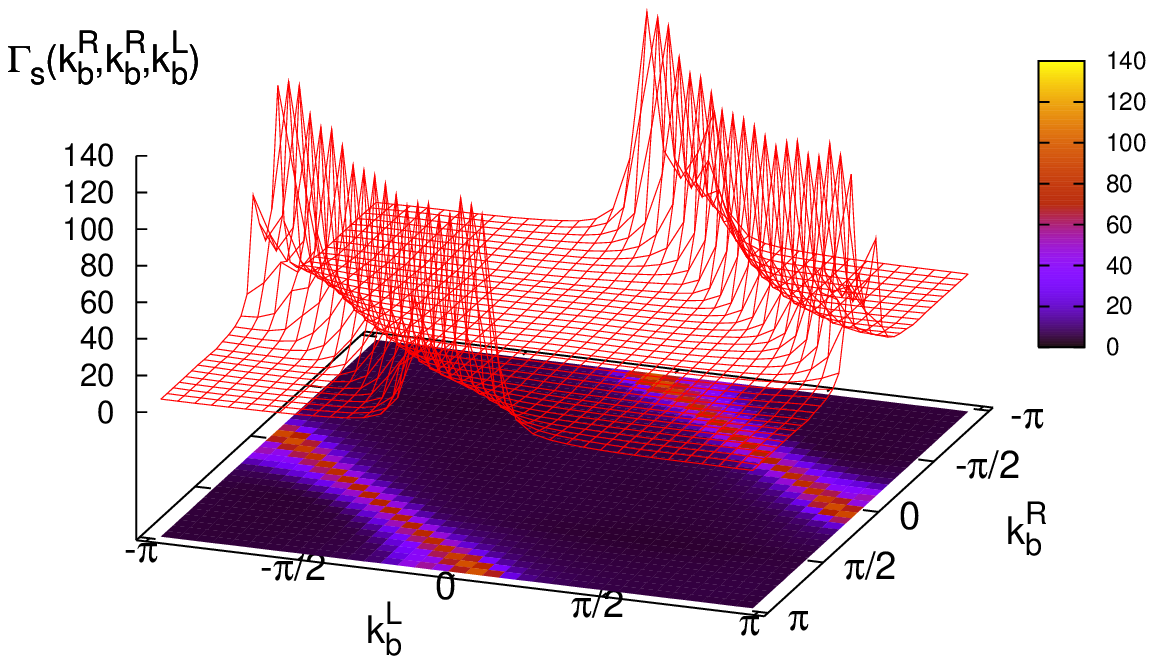,width=7.5cm}
\caption{Same as Figure \ref{frustrated}, but here $\mu=-0.0197$, $U=3.75$. In this case the Fermi surface
intersects the Umklapp surface near $k_b=0$ and $k_b=\pi$.} \label{hdp}
\end{figure}

Upon changing the chemical potential the hot spots mentioned above move along the Fermi surface. When $\mu$ is
increased, the two points in each quadrant eventually merge until the Fermi surface touches the Umklapp surface
at $(\pm\pi/2,\pm\pi/2)$. In figure \ref{edp} we show plots for the vertex function in analogy to figures
\ref{nested} and \ref{frustrated}. Indeed, the vertex function along the Peierls lines in both Umklapp and
Cooper channel exhibits a strong increase in the diagonal region, corroborating the identification of the hot
spots as originating from the points where the Fermi surface intersects the Umklapp surface.  For even larger
values of the chemical potential these points do not exist anymore and thus Umklapp processes will not feed back
into the self energy. Similarly, upon decreasing $\mu$ the eight hot spots move towards the axis and eventually
merge to form four hot spots located at $k_b = 0$ and $k_b = \pi$. Once more the structure of the vertex
function reflects this, as can be seen in figure \ref{hdp}, although for the parameters chosen here the
variation along the Fermi surface is somewhat weaker. For small enough values of $\mu$ the hot spots will again
disappear.

%In general, there are eight hot spots, i.e. two in each quadrant of the BZ. The locations
%of these hot spots neither agree with the scenario put forward by Essler and
%Tsvelik\cite{Ess_02,Ess_05}, nor with the RG calculations by Duprat and Bourbonnais \cite{Dup_01}.
%However, it is in perfect agreement with results obtained for the frustrated two-dimensional case\cite{Roh_05}.

%%% Conclusion

\section{Discussion and Conclusion}

In summary, we have studied a quasi one-dimensional model of coupled chains at and near
half-filling, using an fRG technique and focusing on the appearance of ``hot'' regions
on the Fermi surface. In the presence of perfect nesting, we have found that the whole
Fermi surface is hot. In contrast, in the presence of frustration ($t'_b\neq 0$), isolated
hot spots appear. The mechanisms for the formation of these hot spots is that the effective couplings
(vertex functions) in the various channels become large in an anisotropic manner.
The location of the hot spots corresponds to the intersection of the Umklapp surface with the
Fermi surface.
In general, there are eight hot spots, i.e. two in each quadrant of the Brillouin zone.
Their precise location depends on the doping level and on the ratio $t'_b/t_b$.
These results are perfectly consistent with previous weak-coupling fRG studies of the frustrated two-dimensional
case\cite{Roh_05}.
The location of these hot spots do not agree however with a previous weak-coupling RG study of
a quasi one-dimensional model~[\onlinecite{Dup_01}]. As we have seen, a proper treatment of
Umklapp processes, which are a key ingredient to the mechanism described in the present work, is essential. Because
Ref.[\onlinecite{Dup_01}] was motivated by the strongly metallic regime, it did
not explicitly include these processes in the RG treatment, besides
a mere renormalisation of the forward scattering amplitude.
%AG:
%[DO YOU AGREE w/ THE ABOVE SENTENCE ? FEEL FREE TO PHRASE IT DIFFERENTLY OR MORE MILDLY]

There are naturally also some limitation to our fRG calculation. First,
it is valid only in the weak-coupling regime. Second, the argument that a strongly peaked interaction can
create hot spots on the Fermi surface relies on the low-dimensional properties of the model, the reason being
that with increasing dimension the feedback of the interaction onto the one-electron self-energy via the
two-loop diagram weakens and is eventually washed out. The method does take transverse Umklapp processes into
account, which is essential to the main mechanism and observed features.
This can already be achieved by a much simpler RPA calculation. However, the fRG not only provides an
exact starting point relying on rigorous statements, it also
modifies the RPA results. The critical scales are lower and in contrast to RPA the properties in Cooper and
Umklapp channels are different at equal momentum transfer.

A different location of the hot spots was also found by strong-coupling techniques, such as
the resummation of the expansion in the inter-chain hopping of Ref.~[\onlinecite{Ess_02,Ess_05}] and
the recent chain-DMFT treatment of Ref.~[\onlinecite{Ber_06}]. This is less surprising, and
it is tempting to speculate that the hot spot location may be determined by the regions in
momentum space where the effective couplings are big in the weak coupling limit, while it is
associated with regions in which the inter-chain kinetic energy is small in the strong coupling
limit. Future studies at intermediate coupling are needed in order to elucidate this point and provide
a consistent picture of how the location of the hot spots evolve from weak to strong coupling.

The possibility that electron-electron Umklapp scattering may account for the emergence of hot spots along a
quasi-1d Fermi surface was first suggested by Chaikin in order to explain magic angles in the magnetoresistance
data of Beechgaard salts\cite{Cha_92}. Here we have shown on the basis of a microscopic model and a
functional renormalization group approach how such a
situation may arise. Angular dependent magnetoresistance oscillation experiments only provide
indirect evidence for the formation of hot spots however, through a momentum dependence of the scattering
rate on the Fermi surface which requires theoretical modelisation. Obviously, direct angular-resolved
spectroscopy experiments (e.g photoemission), although very difficult to perform on quasi one-dimensional
organic conductors, would be highly desirable in order to probe these effects experimentally.

%AG: I've suppressed this and displaced the ref to the introduction, do you agree ? cf my email
%It should be stressed that the mechanism at work, namely large and strongly anisotropic
%low-energy scattering processes, is different to a mechanism which involves a more sophisticated description of
%the one-particle band structure, while assuming that the interaction remains isotropic.\cite{Zhe_95}

\acknowledgments
We are grateful to T.~Giamarchi and C.~Bourbonnais for valuable discussions. Support for this work was provided
by CNRS and Ecole Polytechnique, and by the E.U. ``Psi-k $f$-electron'' Network under contract
HPRN-CT-2002-00295.

%%%%%%%%%%%%%%%%%%%%%%%%%% REFERENCES %%%%%%%%%%%%%%%%%%%%%%%%%%%%%%%%%%%%%

\vfill\eject

\end{document}